# Spreadsheet Risk Management in Organisations

Ben G. Rittweger and Eoin Langan
Athlone Institute of Technology,
Athlone, Ireland
Email: benrittweger@hotmail.com, elangan@ait.ie

**ABSTRACT**

*The paper examines in the context of financial reporting, the controls that organisations have in place to manage spreadsheet risk and errors. There has been widespread research conducted in this area, both in Ireland and internationally. This paper describes a study involving 19 participants (2 case studies and 17 by survey) from Ireland. Three areas are examined; firstly, the extent of spreadsheet usage, secondly, the level of complexity employed in spreadsheets, and finally, the controls in place regarding spreadsheets. The findings support previous findings of Panko (1998), that errors occur frequently in spreadsheets and that there is little or unenforced controls employed, however this research finds that attitudes are changing with regard to spreadsheet risk and that one organisation is implementing a comprehensive project regarding policies on the development and control of spreadsheets. Further research could be undertaken in the future to examine the development of a "best practice model" both for the reduction in errors and to minimise the risk in spreadsheet usage.*

## 1 INTRODUCTION

The impact that spreadsheets have made to accounting has been transformational. They are used in many different ways and they vary in degrees of complexity. They are used for complex financial models, financial reporting, data storage and analysis, presentation of results and simple one-off calculation. With such widespread use, it is important that the results can be relied upon. There is much academic research published on spreadsheet errors and the risk associated with spreadsheets. Although many spreadsheets are used for small "scratchpad" applications, they are also used to develop many large applications. In general spreadsheet errors tend to occur in only a few percent of cells, therefore in large spreadsheets it is not if there is an error but how many errors are present, Panko (1998).

Many organisations are becoming more aware of the potential problems with spreadsheets as a result of Sarbanes-Oxley (2002) "SOX" and the Financial Services Authority (UK) emphasis on systems and controls. In the current market environment it is vital that companies ensure their operations are under control. Previously companies have often focused risk management resources on financial reporting processes, however it has become apparent that control over operational and decision making processes is just as important, as these processes are often highly spreadsheet dependent. There have been high profile cases caused by malfunctioning spreadsheets such as the Allfirst/AIB scandal. John Rusnack doctored some of his spreadsheets to present a different trading position, Butler (2009).

**1.2 The extent of spreadsheet use within in organisations**
There is substantial research to confirm that spreadsheets are used in almost all businesses for financial reporting. According to Panko (1998):

- Financial intelligence firm CODA reports that 95% of U.S. firms use spreadsheets for financial reporting according to its experience.
- In 2004, RevenueRecognition.com (now Softtrax.com), had the International Data Corporation interview 118 business leaders. IDC found that 85% were using spreadsheets in reporting and forecasting and that 85% were using spreadsheets for budgeting and forecasting.
- In 2003, The Hacket Group surveyed mid-size companies. It found that 47% of companies use stand-alone spreadsheets for planning and budgeting.

**1.3 Current research findings**
Panko (1998), summarised the results of seven field audits in which operational spreadsheets were examined, typically by an outsider to the organisation. His results show that 94% of spreadsheets have errors and that the average cell error rate (the ratio of cells with errors to all cells with formulas) is 5.2%. Table 1 summarises the data behind these estimates. In total, 88 spreadsheets are represented in the table. For all 88, the weighted average percentage of spreadsheets with errors is 94%. Data on cell error rates were available on 43 of these spreadsheets, and the weighted average for this sample is 5.2%.

Table 1: Spreadsheet error rates (1998)

| Source | Number Audited | Percent with Errors | Cell Error Rate (% of cells) | Comment |
|---|---|---|---|---|
| 1. Coopers and Lybrand (1997) | 23 | 91 | n/a | Off by at least 5% |
| 2. KPMG (1998) | 22 | 91 | n/a | Only significant errors |
| 3. Hicks (1995) | 1 | 100 | 1.2 | One omission error would have caused an error of more than $1bn. |
| 4. Lukasic (1998) | 2 | 100 | 2.2 | In model 2, the investment model was overstated by 16%, Quite serious |
| 5. Butler (2000) | 7 | 86 | 0.4 | Only errors large enough to require additional tax payments ** |
| 6. Clermont (2002) | 3 | 100 | 3.0 | Computed on the basis of non-empty cells |
| 7. Lawrence and Lee (2004) | 30 | 100 | 6.9 | 30 most finically significant SS's audited by Mercer Finance & Risk Consulting in previous year |
| Average | 88 | 94* | 5.2** | |

*Weighted average of 88 spreadsheets in sources 1-7.*
**Weighted average of 43 spreadsheets in sources 3-7.*

If Hicks (1995) is examined as an example in Table 1, this was an audit on a large capital budgeting spreadsheet with 3,856 cells. It was a three person code by code inspection, errors were found in 1.2% of all lines of code, but one error if not detected would have cost over $1bn. This demonstrates the complexity of spreadsheets and how effective a full code inspection was to ensuring the accuracy of the information. Hendry and Green (1994) did a similar ethnographic study with ten spreadsheet developers. They added a phase in which the developer walked through a spreadsheet with the interviewer. They found that the developers had a hard time in explaining parts of the spreadsheet that they themselves had built, demonstrating the informal approach that developers have regarding spreadsheet development.

At present there is no "silver bullet" solution in place for detecting spreadsheet errors in an efficient and timely manner Panko (1998). Panko further explains that to control errors you must understand them. There are a number of error taxonomies that have been put forward to highlight the differences between different types of errors. There is no one best error taxonomy as each one has its own strengths and weaknesses. One of the most widely used taxonomies is the mistakes/slips/lapses taxonomy created by Norman (1984) and Reason & Mycielska (1982). They look at the distinction between mistakes and slips or lapses. Mistakes are errors in intention, if something is done that is inherently incorrect, this is a mistake. Lapses and slips are errors in executing a right intention. This taxonomy is designed to explain thinking and execution but is used in observation and therefore it is questionable, if it could be applied to finished spreadsheets.

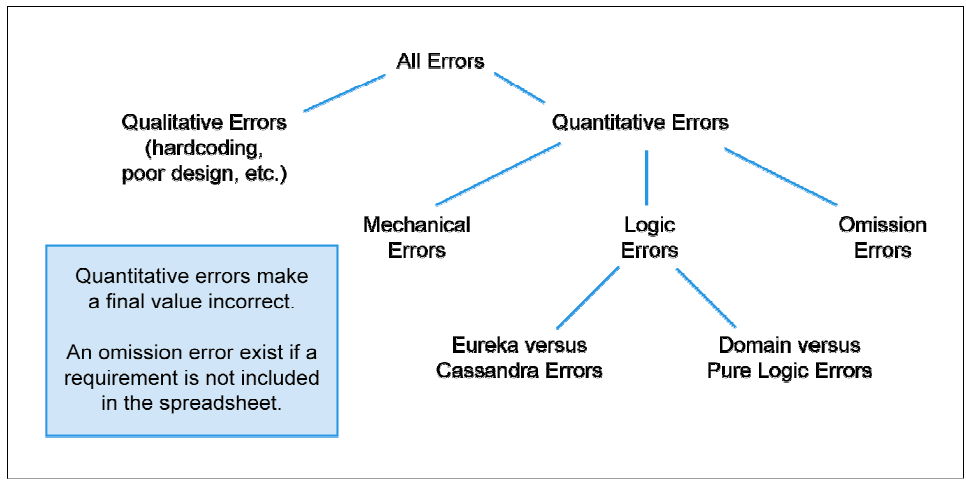

*Figure 1: Panko and Halverson 1996 Taxonomy of Development and Testing    Error Types*

Panko and Halverson (1996) created a taxonomy with several distinctions as seen in figure 1. Firstly, it distinguishes between quantitative errors which give a wrong number immediately and qualitative errors which are likely to lead to wrong numbers later. Secondly, based on Allwood's (1984) work in mathematics, the taxonomy distinguishes between mechanical errors, logical errors in creating formulas and omission errors; which is the result of leaving something out of a model. In relation to logic errors, it distinguishes between "Eureka" errors (Lorge and Solomon, 1955), which are easily proven, and "Cassandra" errors, which are difficult to prove even when they are detected.

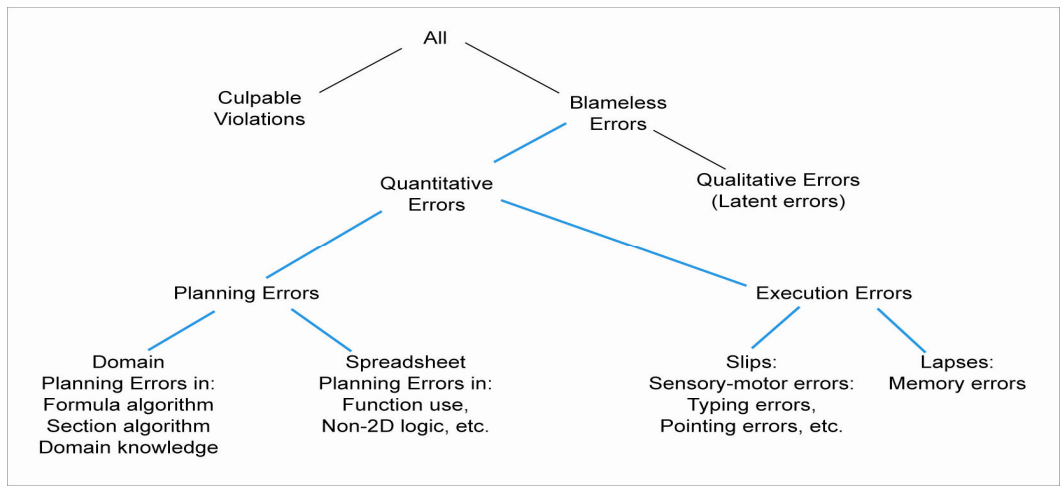

*Figure 2: Revised Taxonomy of Spreadsheet Errors*

Panko and Halverson (2010) revised the taxonomy in figure 2, although the original taxonomy had been fairly well validated by experiments, some limitations have become obvious over time. The taxonomy did have an error type dimension and a spreadsheet life cycle perspective, however it did not flesh out the lifecycle dimension. Errors that occurred during the initial analysis and requirements were not examined and they did not study the ongoing use and were not aware of overwriting errors, in which a user overwrites a formula with a number. Omission errors were focused on because they were the subject of earlier human error research. Omission of a requirement is only one type of noncompliance. Finally, the taxonomy did not recognise the important distinction between sensory-motor slips and memory lapses.

The requirements expected by finance personnel to prepare and analyse large amounts of information within very tight time deadlines is not uncommon. This could be argued as a factor that induces errors. Spreadsheets that carry out hundreds and in some cases thousands of calculations are inherently complex and can give rise to errors. Some inaccuracies are culpable though, Panko (2007) has highlighted 3 examples:

- Violations and wilful departures from company policy (if in place).
- Puffery or window-dressing a position using incorrect facts or assumptions.
- Building spreadsheets when the developer lacks sufficient experience.

These examples raise further questions, it maybe difficult to determine if a user deviates. While the developer might be competent in advanced Excel, if there is no clear policy within the organisation he/she cannot be considered an expert or an inexpert.

**1.4 Controls and solutions to control spreadsheet risk**
Galletta & Hufnagel (1992), surveyed 107 Management Information Systems "MIS" executives using a mail questionnaire that focused broadly on policies for controlling End User Development "EUD". For restrictions on EUD, 23% said they had rules and 58% said they had guidelines. Yet even for corporations with rules and guidelines, the respondents said that there was full compliance only 27% of the time. When asked about requirements for post-development audits, only 15% said they had rules, and another 34% said they had guidelines. The respondents said that where they had rules and guidelines for post-development audits, there was full compliance only 10% of the time, and guidance was completely ignored 41% of the time. All of these statistics are only for EUD that the MIS managers are aware of.

Panko (2007) examines how logic inspection can be done, adopting Fagan's (1976, 1986) code inspection methodology. Logic inspection (just look at everything), is the most suitable method according to Panko (2007). This is called code inspection in software development, which inspects a program module, line by line, looking for errors. In logic inspection for spreadsheets, inspectors examine all formula cells. This approach can be applied in the design phase before coding begins. Panko (2007) identifies multiple types of testing:

- Test during development and separately through each phase.
- Requirements testing: Many errors are introduced before coding ever begins.
- Unit testing: After a developer has finished a module and implemented their own self check, the module must be subjected to unit testing.
- Integration testing: After modules are tested, they are integrated into larger units. Usually several stages of integration are needed, each with its own techniques for unit testing. According to Fagan (1976, 1986) testing by one individual will only catch 50% to 60% of errors. Team inspection can raise the detection rate to 80%.
- Agile development methods: It is assumed that a traditional software development life cycle "SDLC" model is employed. Spreadsheet development maybe done in non-traditional ways, especially agile methods.
- Eyeballing: One testing technique is looking over the spreadsheet for reasonableness or having a colleague check a spreadsheet. There is no evidence that eyeballing reduces error rates.
- Error scanning software: Excel 2003 has a built in error checking tool under the tools menu. This is simple but limited. Error checking software products such as SpACE, Comply XL, Cluster Seven and Acuate which can tend to assist in locating errors within complex spreadsheets. However, they would never be able to detect a quantitative error such as an omission.
- Auditing: In auditing, an auditor does not examine everything; they ask questions whose answers may indicate problems. Auditing will only perform spot checks and the goal of auditing is to detect indications of problems and not to reduce errors.

Butler (1998) contends that in the case of AIB, that if the software SpACE was employed, it would have highlighted the existence of the links in the pricing spreadsheet and explained where they went to enable follow-up. Panko (1998), states that EUD in spreadsheets is so casual as developers and organisations are overconfident about the accuracy of spreadsheets despite the large amount of information to the contrary. Panko furthermore states that overconfidence is corrosive because it tends to blind people to the need for reducing risk and when errors are detected we become convinced of our ability to detect errors. Chambers and Hamill (2008) undertook a case study within a mid sized international bank on End User Applications risk and concluded that risk was poorly understood and rarely controlled in the bank. However, McGeady & McGouran (2008), described the development and implementation of an effective End User Computing "EUC" policy within AIB Capital

Markets, they concluded that EUC has arrived and is here to stay and with proper controls in place EUC can be a valuable asset to an organisation.

**1.5 Rationale**
This research was undertaken in part fulfilment of an online undergraduate honours degree in Applied Accounting. The researcher is a mature student returning to education with many years in senior accounting roles and is an Associate Member of The Chartered Institute of Management Accountants. The rationale for this research derives from the researcher having a personal interest in this area within his current role, in the context that spreadsheet models have been developed as management tools to aid the decision making process and that they are used extensively in financial reporting. The research aims to leverage both the international and Irish research on the issues surrounding spreadsheet error and risk. The aim of the current research is to investigate within the context of financial reporting, the controls that organisations have in place to control spreadsheet risk and errors. The research had a clear focus with three core research objectives:

1. To investigate the level and the extent that spreadsheets are used in financial reporting.
2. To explore the level of complexity that is employed in the use of spreadsheets and examine the impact of errors identified.
3. To identify the controls if any and the possible solutions to control spreadsheet risk and errors.

The remainder of this paper will examine the background of spreadsheet error and risk, describe the methodology employed in the research, provide discussion of the results and a summary conclusion.

## 2 METHODOLOGY

Due to the explanatory nature of the research, a mixed method approach was adopted using a case study and a survey strategy approach as the best option for the research strategy; it was envisaged that the common themes identified from the case study would be reinforced from the findings in the survey.

Two data collection methods were employed, semi-structured interviews in the context of the case studies and a questionnaire in the context of a survey strategy.

**2.2 Semi-structured interview**
Certain information could only be obtained by direct conversation with individuals who would have access to sensitive information. Interviewees were selected on the basis that they were willing to participate. Three organisations had initially been chosen and had agreed to participate in the research. Two of the participants are employed in indigenous organisations, one is a finance manager in the retail banking sector and the other is an IT manager, who works closely in conjunction with the finance department in the pension/life assurance sector; both are quoted on the Irish Stock Exchange. The third interviewee at short notice was unable to commit.

Prior to conducting the interviews, it was explained to each of the participants that any information provided by them in the interview would be treated in the strictest of confidence and that anonymity would be preserved to prevent individuals from being identified in the research findings. The two interviewees are identified as P1 and P2 in the analysis of the semi-structured interviews. The interviews were recorded with participants' approval. A neutral location was chosen for the two interviews. Prior to conducting the interviews, participants were given an outline of the research being undertaken.

## 2.3 Questionnaire

The use of surveymonkey.com as the researcher's questionnaire provided a number of additional benefits when compared to other types of questionnaires. The researcher could design a questionnaire with a high quality appearance with relative simplicity in designing the format. Surveymonkey.com automatically creates a link to your survey after you have completed the design, which you can include in your email cover letter. The respondent can click on the link and answer the questionnaire with all answers automatically recorded, these can be analysed at a later stage. In addition to enhance confidentiality of the information provided, all pages can be Secure Sockets Layer (SSL) encrypted. The questionnaire was designed in a format similar to the questions employed in the semi-structured interviews.

## 2.4 Sample population

Management finance professionals across a number of industries were targeted in the research. The sampling frame was constructed from using finance professionals from the Linkedin website. An email was sent to the sample population in April 2010 with a link to the questionnaire, inviting them to participate in the research. It outlined who the researcher was, the nature of the research and requesting their participation by completing the questionnaire. The link on the email ensured that confidentiality of users was preserved. In total 30 questionnaires were distributed, of which 17 (57%) responses were returned, which were all usable. The demographic profile highlighted that 88% of the population work in four sectors: logistics, manufacturing, services and financial services.

## 3 FINDINGS

This analysis of the results addresses the three key objectives of the research:

1. To investigate the level and the extent that spreadsheets are used in financial reporting.
2. To explore the level of complexity that is employed in the use of spreadsheets and examine the impact of errors identified.
3. To identify the controls if any and the possible solutions to control spreadsheet risk and errors.

**3.2 Objective one**: **To investigate the level and the extent that spreadsheets are used in financial reporting.**
Spreadsheet usage seems to be high and diverse across the two organisations in the case studies, in the context of financial reporting, both interviewees confirmed that Microsoft Excel was the only spreadsheet package used in their organisations, respectively. P1 stated that *"Spreadsheets are heavily used in financial reporting. Some final reports, though not in Excel format are derived from Excel spreadsheets",* P2 stated that *"They are used across the board, the main use for them is that they are so easy to use and to get up and running and you become effective with them in a short space of time".* Both organisations use data warehousing models to store data. P2 explained that there are standard reports available in data warehouse systems and in addition there is constant demand for one-off reports which is satisfied by allowing end users interrogate the data warehouse. The data is then formatted in Excel using pivot tables and other formulas, resulting in the desired information. Both interviewees confirmed that their front office personnel do not use spreadsheets for key operational activities, only mid office and back office personnel use spreadsheets that contain data from key systems or data warehousing models. Both interviewees confirmed that spreadsheet risk was a moderate to high issue within their organisations and this view was shared by senior management.

Results from questions 1 and 4 from the questionnaire triangulate the findings identified in the semi-structured interviews regarding the level and the extent that spreadsheets are used in financial reporting.

Question 1
Question 1 (figure 3) asked participants to rate the context that spreadsheets are used in their organisations. Spreadsheets are used 65% all of the time in financial reporting and management accounting. It could be derived that although spreadsheets are used extensively in a financial reporting context, it seems that they are relied on some of the time in other functions of an organisation such as, decision making 53% and operations 47%, respectively.

| In general in what context are spreadsheets used in your organisation? | | | | | |
|---|---|---|---|---|---|
| Answer Options | Never | Sometimes | All of the time | Rating Average | Response Count |
| Financial reporting | 1 | 4 | 11 | 2.63 | 16 |
| Operations | 0 | 6 | 9 | 2.60 | 15 |
| Management accounting | 0 | 6 | 11 | 2.65 | 17 |
| Decision making | 1 | 6 | 8 | 2.47 | 15 |
| Can you specify the sector your business is most closely related to? | | | | | 17 |
| | | | | answered question | 17 |
| | | | | skipped question | 0 |

*Figure 3: Spreadsheets Usage in Organisations*

Question 4
16 out of 17 participants answered question 4 regarding spreadsheet risk. 75% regard spreadsheet risk as a moderate risk with 12.5% regarding spreadsheet risk as a high risk, in the semi-structured interviews both participants confirmed that spreadsheet risk was a moderate to high risk issue.

**3.3 Objective two: To explore the level of complexity that is employed in the use of spreadsheets and examine the impact of errors identified.**
P1 explained that core spreadsheets used in their business unit contained basic formulas using warehouse or core systems data to verify balances as a control check. The complexity of spreadsheets is very low and the number of formulas is not more than ten in most cases, although there would be a number of links between different spreadsheets. Internal audit conducts a sample testing on a quarterly basis to test the accuracy and validity of core spreadsheets, this is part of the process in ensuring compliance with SOX. P2 stated that in the operations area of the business, *"Qualified mathematicians would use Excel as a programming tool to develop complex models in the core areas of the business",* although this is an operational function, the researcher contends that it is within the context of answering the objective, as the organisation sells financial products. P2 confirmed that there are highly complex formulas involved in these models. P2 further confirmed that internal audit examines a sample of spreadsheets on a periodic basis.

Results from questions 6 and 8 from the questionnaire triangulate the findings identified in the semi-structured interviews regarding the level of complexity that is employed in the use of spreadsheets and the impact of errors identified.

Question 6
16 out of 17 participants answered question 6 (figure 4) regarding complexity, this question was trying to gauge the level of complexity employed in spreadsheets by determining the number of formulas employed. 75% of respondents indicated that there were <100 formulas employed in a complex spreadsheet.

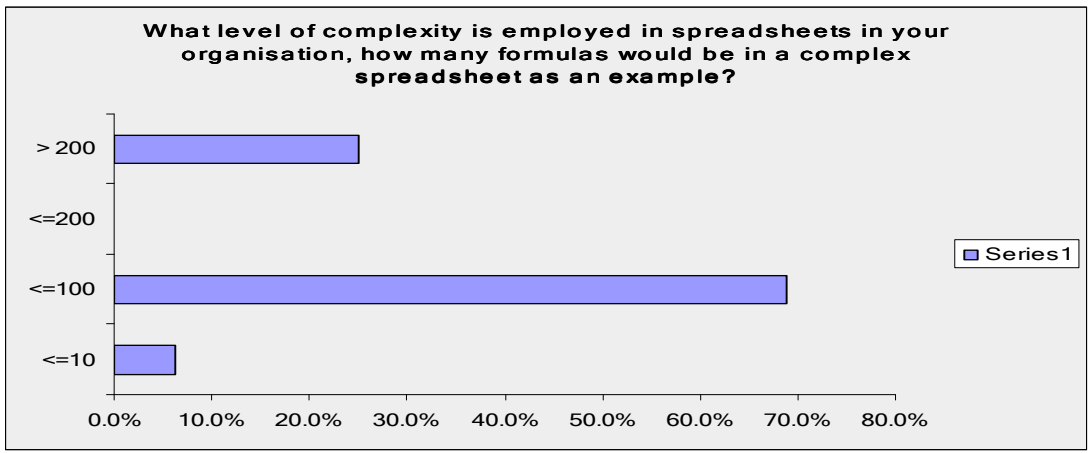

*Figure 4: Complexity*

Question 8
12 participants answered question 8 (figure 5) regarding spreadsheet errors, which equates to 71% of the sample population. 11 participants stated that errors detected had a value of less than €10k and one interestingly had a value between €10k and €50k.

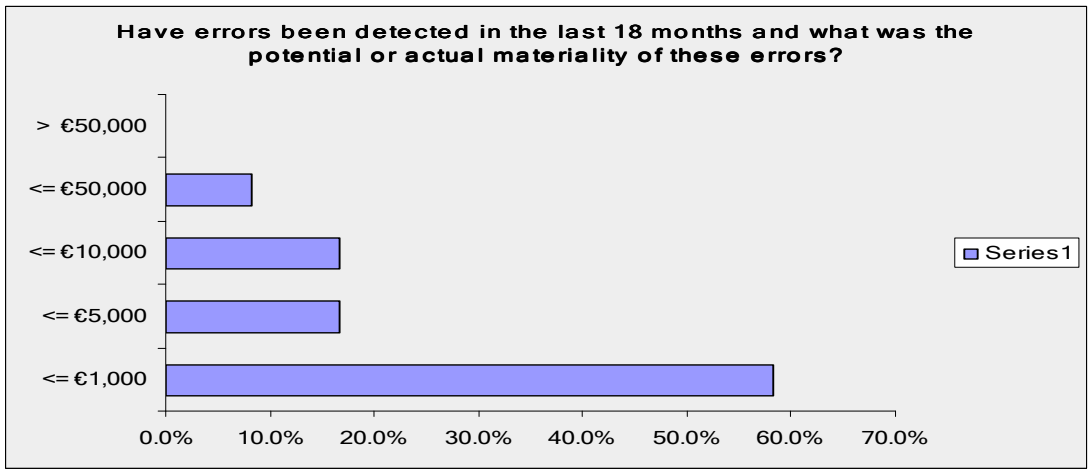

*Figure 5: Spreadsheet Errors*

**3.4 Objective three**: **To identify the controls if any and the possible solutions to control spreadsheet risk.**

Both interviewees confirmed that there were controls in place with key spreadsheets having restricted access on a local area network "LAN" with password protection, P1 stating that *"absolutely, under SOX there would be a lot of controls around how information is recorded and kept, because we have a listing on the American Stock Exchange, we have to abide by the SOX rules, which resulted in a huge exercise around controls and ownership called End User Computing, this is were you have to set out what spreadsheets are used and how they are controlled"*. P2 stated that *"It has been getting a head of steam in the last year, in the past it has been managed in terms of coding standards in how spreadsheets should be developed and ring fence them on the LAN in a certain place so they can be monitored, I don't know how effective this was though in issuing coding standards and that people are using them"*. P2 further explained that a project was in the pipeline regarding introducing software to test spreadsheets errors and control risk. A consultancy firm had undertaken a pilot project recently, with a group wide rollout to be undertaken later this year at a cost of €600k. P2 confirmed that this consultancy was also used by a different business segment in P1's organisation, although P1 was not aware of this at the time of the interview.

Results from questions 7, 9 and 10 from the questionnaire triangulate the findings identified in the semi-structured interviews regarding the controls in place and the possible solutions to control spreadsheet risk.

Question 7
All participants in this survey responded to question 7 (figure 6) regarding control policy, with 59% stating that there were controls in place regarding access, control and validation functions. This highlights that there are organisations that do not have in place basic access controls. From his own experience the researcher contends that small organisations may have informal policies where there are only one or two finance professionals employed.

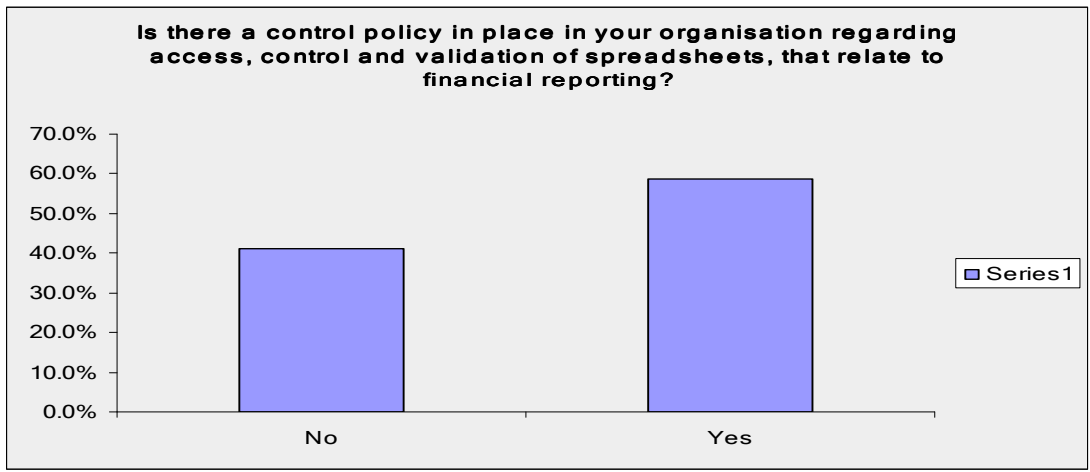

*Figure 6: Control Policy*

Question 9 & Question 10
16 participants answered question 9 regarding auditing, with 81% stating that there is no software tools in place to audit spreadsheets for errors. 17 participants answered question 10 regarding developing spreadsheet policies in the future, 64% of respondents stating no, 24% stating maybe and 12% stating yes.

## 4 CONCLUSION

The researcher contends that the aim of investigating in the context of financial reporting, the controls that organisations have in place to control spreadsheet risk and errors has been achieved. A number of themes were identified when answering the objectives, these were triangulated with the findings in the questionnaire, they are as follows:

1. Spreadsheet usage is high and is used extensively in financial reporting and spreadsheet risk is considered to be an important issue within organisations. This confirms previous research findings of Panko (1998) and confirms that spreadsheet usage in Ireland is comparable to international research. Furthermore the findings show that there is a high reliance on spreadsheets within a financial reporting context and that they are used extensively. Finally, spreadsheet risk is considered to be an important issue within an organisation.

2. Internal audit have conducted a sample of core/complex spreadsheets in an organisation to test for accuracy and completeness. The researcher contends that, although internal audit have checked spreadsheets to identify errors, it is unclear what methods or software they have employed to achieve this.

3. Controls are somewhat in place regarding core spreadsheets with access restricted within LAN's. Although policies maybe in place in organisation, they are not rigorously adopted in most cases, however with internal controls becoming more part of senior management responsibilities, it is changing attitudes within organisations.

### 4.1 Recommendations
The researcher contends that organisations need to adopt and enforce policies to control the risk of errors occurring. The biggest risk which Panko (1998), has identified is the omission error; leaving something out of a model. Thorne (2009) confirms that there is audit software and tools to detect spreadsheet errors, however the impact is unknown regarding how effective they are in reducing spreadsheet errors. These errors can only be detected if proper SDLC's are adopted in organisations regarding core spreadsheets, therefore the researcher is of the opinion that it is impractical to have a SDLC for every spreadsheet.

Further research could be conducted that would include other professionals in an organisation such as IT and operations as well as finance professionals; to gain a more in depth understanding of how organisations manage spreadsheet risk. A questionnaire using a larger sample could be applied as it possesses an anonymous trait, which may result in participants disclosing sensitive information. Further research should consider the development of a best practice model both for the reduction in errors and to minimise the risks.